\documentclass{pasj00}

\DeclareAbbreviation\AAHam{Astron. Abh. Hamburg. Sternw.}
\DeclareAbbreviation\AARv{Astron. Astrophys. Rev.}
\DeclareAbbreviation\an{Astron. Nachr.}
\DeclareAbbreviation\AcA{Acta Astron.}
\DeclareAbbreviation\Afz{Astrofizika}
\DeclareAbbreviation\AnTok{Tokyo Astron. Obs. Annals, Sec. Ser.}
\DeclareAbbreviation\Ap{Astrophysics}
\DeclareAbbreviation\ARep{Astron. Rep.}
\DeclareAbbreviation\ATel{Astronomer's Telegram}
\DeclareAbbreviation\ATsir{Astron. Tsirk.}
\DeclareAbbreviation\AcApS{Acta Astrophys. Sinica}
\DeclareAbbreviation\AstL{Astron. Lett.}
\DeclareAbbreviation\BaltA{Baltic Astron.}
\DeclareAbbreviation\BASI{Bull. Astron. Soc. India}
\DeclareAbbreviation\BeSN{Be Newslett.}
\DeclareAbbreviation\GCN{GRB Coord. Netw. Circ.}
\DeclareAbbreviation\ibvs{Inf. Bull. Variable Stars}
\DeclareAbbreviation\JAD{J. Astron. Data}
\DeclareAbbreviation\JAVSO{J. American Assoc. Variable Star Obs.}
\DeclareAbbreviation\JBAA{J. British Astron. Assoc.}
\DeclareAbbreviation\JO{Journal des Observateurs}
\DeclareAbbreviation\LowOB{Lowell Obs. Bull.}
\DeclareAbbreviation\MitVS{Mitteil. Ver\"{a}nderl. Sterne}
\DeclareAbbreviation\MmSAI{Mem. Soc. Astron. Ita.}
\DeclareAbbreviation\Msngr{Messenger}
\DeclareAbbreviation\NewA{New Astron.}
\DeclareAbbreviation\NewAR{New Astron. Rev.}
\DeclareAbbreviation\NZJS{New Zealand Journal of Science}
\DeclareAbbreviation\OAP{Odessa Astron. Publ.}
\DeclareAbbreviation\Obs{Observatory}
\DeclareAbbreviation\PASA{Publ. Astron. Soc. Australia}
\DeclareAbbreviation\PAZh{Pis'ma AZh}
\DeclareAbbreviation\PhR{Phys. Rep.}
\DeclareAbbreviation\PVSS{Publ. Variable Stars Sect. R. Astron. Soc. New Zealand}
\DeclareAbbreviation\PZ{Perem. Zvezdy}
\DeclareAbbreviation\PZP{Perem. Zvezdy Pril.}
\DeclareAbbreviation\QJRAS{QJRAS}
\DeclareAbbreviation\RMxAA{Rev. Mexicana Astron. Astrof.}
\DeclareAbbreviation\RvMA{Reviews of Modern Astron.}
\DeclareAbbreviation\Sci{Science}
\DeclareAbbreviation\SvA{Soviet Astronomy}
\DeclareAbbreviation\SvAL{Soviet Astronomy Letters}
\DeclareAbbreviation\VeSon{Ver\"{o}ff. Sternw. Sonneberg}
\DeclareAbbreviation\VSOLJBul{VSOLJ Variable Star Bull.}
\DeclareAbbreviation\yCat{VizieR Online Data Catalog}
\DeclareAbbreviation\ZA{Z. Astrophys.}

\def\PublisherCambridge{Cambridge: Cambridge University Press}

\def\PublisherUAP{Tokyo: Universal Academy Press}

\begin{document}
\SetRunningHead{T. Ohshima et al.}{Discovery of Negative Superhumps in ER UMa}

\Received{2011/03/09}
\Accepted{2012/06/20}

\title{Discovery of Negative Superhumps during a Superoutburst
of January 2011 in ER Ursae Majoris}

\author{Tomohito Ohshima,\altaffilmark{1} Taichi Kato,\altaffilmark{1}  Elena Pavlenko,\altaffilmark{2}   Hiroshi Itoh,\altaffilmark{3} Enrique de Miguel,\altaffilmark{4,5} Thomas Krajci,\altaffilmark{6} Hidehiko Akazawa,\altaffilmark{7} Kazuhiko Shiokawa,\altaffilmark{3} William Stein,\altaffilmark{8}Alex Baklanov,\altaffilmark{2} Denis Samsonov,\altaffilmark{2} Oksana Antonyuk,\altaffilmark{2} Maksim Andreev,\altaffilmark{9} Kazuyoshi Imamura,\altaffilmark{7} Franz-Josef Hambsch,\altaffilmark{10} Hiroyuki Maehara,\altaffilmark{11} Javier Ruiz,\altaffilmark{12} Shin'ichi Nakagawa,\altaffilmark{13} Kiyoshi Kasai,\altaffilmark{3} Boyd Boitnott,\altaffilmark{8} Jani Virtanen,\altaffilmark{14} Ian Miller,\altaffilmark{8}}
\altaffiltext{1}{Department of Astronomy, Kyoto University, Kyoto 606-8502, Japan}
\altaffiltext{2}{Crimean Astrophysical Observatory, 98409, Nauchny, Crimea, Ukraine}
\altaffiltext{3}{Variable Star Observes League in Japan(VSOLJ)}
\altaffiltext{4}{Departamento de Fisica Aplicada, Facultad de Ciencias Experimentales, Universidad de Huelva, 21071 Huelva, Spain}
\altaffiltext{5}{Center for Backyard Astrophysics, Observatorio del CIECEM, Parque Dunar, Matalasca\~nas, 21760 Almonte, Huelva, Spain}
\altaffiltext{6}{Center for Backyard Astrophysics (New Mexico), PO Box 1351, Cloudcroft, NM 83117, USA}
\altaffiltext{7}{Department of Biosphere-Geosphere System Science, Faculty of Informatics, Okayama University of Science, 1-1 Ridai-cho, Okayama, Okayama 700-0005, Japan}
\altaffiltext{8}{American Association of Variable Star Observers(AAVSO)}
\altaffiltext{9}{Institute  of Astronomy, Russian Academy of Sciences, 361605 Peak Terskol, Kabardino-Balkaria, Russia}
\altaffiltext{10}{Vereniging Voor Sterrnkunde (VVS), Oude Bleken 12, 2400 Mol, Belgium}
\altaffiltext{11}{Kwasan Observatory, Kyoto University, Yamashina-ku, Kyoto, 607-8471}
\altaffiltext{12}{Observatory of Cantabria, Centro de Investigaci\'on del Medio Ambiete (CIMA) Instituto de F\'isica de Cantabria (IFCA), Agrupaci\'on Astron\'omica C\'antabra (AAC), Ctra. de Rocamundo s/n, Valderredible, Cantabria, Spain}
\altaffiltext{13}{Osaka Kyoiku University, 4-698-1 Asahigaoka, Kashiwara, Osaka, 582-8582}
\altaffiltext{14}{Ollilantie 98, 84880 Ylivieska, Finland}


\maketitle

\begin{abstract}
 We report on a discovery of "negative" superhumps during the 2011
January superoutburst of ER UMa. During the superoutburst which started
on 2011 January 16, we detected negative superhumps having a period of
0.062242(9) d, shorter than the orbital period by 2.2\%. No evidence 
of positive superhumps was detected during this observation.
This finding indicates that the disk exhibited retrograde precession 
during this superoutburst, contrary to all other known cases of 
superoutbursts.  The duration of this superoutburst was shorter than 
those of ordinary superoutbursts and the intervals of normal outbursts 
were longer than ordinary ones.  We suggest a possibility that
such unusual outburst properties are likely a result of the disk tilt, 
which is supposed to be a cause of negative superhumps:
the tilted disk could prevent the disk from being filled with materials 
in the outmost region which is supposed to be responsible for 
long-duration superoutbursts in ER UMa-type dwarf novae.
The discovery signifies the importance of the classical
prograde precession in sustaining long-duration superoutbursts. 
Furthermore, the presence of pronounced negative superhumps in this
system with a high mass-transfer rate favors the hypothesis that
hydrodynamical lift is the cause of the disk tilt.  
\end{abstract}

\section{Introduction}

 SU UMa-type dwarf nova is a subgroup of dwarf novae (For reviews, 
\cite{war95book}) and characterized
by the presence of long-lasting, bright outbursts called superoutbursts,
in addition to ordinary dwarf-nova outbursts. During these superoutbursts,
variations with periods a few percents longer than the orbital period, called 
positive (ordinary) superhumps, are ubiquitously observed, and are considered
as the defining characteristics of SU UMa-type dwarf novae 
\citep{vog80suumastars}.

 The cause of this phenomenon is widely believed to be a dynamical
prograde precession of the elongated disk, which is believed to be
formed by a the tidally driven eccentric instability of the disk
excited by the 3:1 resonance to the orbital motion of
the secondary \citep{whi88tidal}. This tidal instability at 
the radius of the 3:1 resonance is generally
considered to enable the disk to expand than in ordinary outbursts
to produce long, bright superoutbursts \citep{osa89suuma}.

 This picture naturally explains the phenomenon that positive 
superhumps with period longer than the orbital period are 
always observed during superoutbursts, and the presence of
positive superhumps is a logical consequence of this picture.
 Indeed, not a single definite exception against the ubiquitous
presence of positive superhumps during superoutbursts has been
reported in the nearly 40 yrs history since the discovery of 
positive superhumps.

 On the other hand, superhumps shorter than the orbital periods 
(negative superhumps) are also reported (\cite{uda88ttari}; 
\cite{har95v503cyg};\cite{rin12v378peg}).

Unlike positive superhumps, negative superhumps are
exclusively detected in the hot, thermally stable disk in
novalike CVs, and during quiescence and some normal outbursts in
a small number of dwarf novae including SU-UMa type dwarf novae
(\cite{pat95v1159ori};\cite{gao99erumaSH};\cite{pav10mndra}).
Negative superhumps are detected in V344 Lyr except during
the superoutburst (\cite{sti10v344lyr}, \cite{woo11v344lyr}. 
 As to negative superhumps in superoutburst stage, a few cases
have been reported (\cite{pat95v1159ori}, \cite{can12v1504cyg},
)
but otherwise yet not recorded during superoutbursts.

 The origin of negative superhumps
has not been well-understood. They are usually considered as a result
of some kind of retrograde precession in the accretion disk.
There are several suggestions that the torque by the secondary 
on the tilted or
warped disk produces a retrograde precession. Especially \citet{mon09}
suggested that this retrograde precession in the disk is due to the same
tidal force as the Moon on the retrogradely processing Earth.
However, the mechanisms
for producing a tilt or a warp are still controversial (\cite{woo00SH};
\cite{mur02warpeddisk}). 

 ER UMa is a member of SU UMa-type dwarf novae, whose intervals of
superoutburst (supercycle) are very short \citep{kat95eruma}. This object
is known as the prototype of a subgroup,
``ER UMa type'' among SU-UMa type stars (e.g. \cite{rob95eruma}; for a
review see \cite{kat99erumareview}). 
 In terms of the presence of positive superhumps during superoutbursts,
ER UMa shares common properties with other SU UMa-type dwarf novae 
\citep{kat03erumaSH}, although peculiar aspects of its behavior have been
reported (\cite{kat96erumaSH}; \cite{kat03erumaSH}).
Although \citet{gao99erumaSH} and \citet{km10eruma} suggested on 
the presence of negative
superhumps during quiescence and a normal outburst, only positive 
superhumps were observed during the following superoutburst.

 In this \textit{Letter}, we report on the discovery of ``negative'' superhumps
during the 2011 January superoutburst of ER UMa.

\section{Observations and Result}

\begin{figure}[!htb]
\begin{center}
\FigureFile(70mm,157.5mm){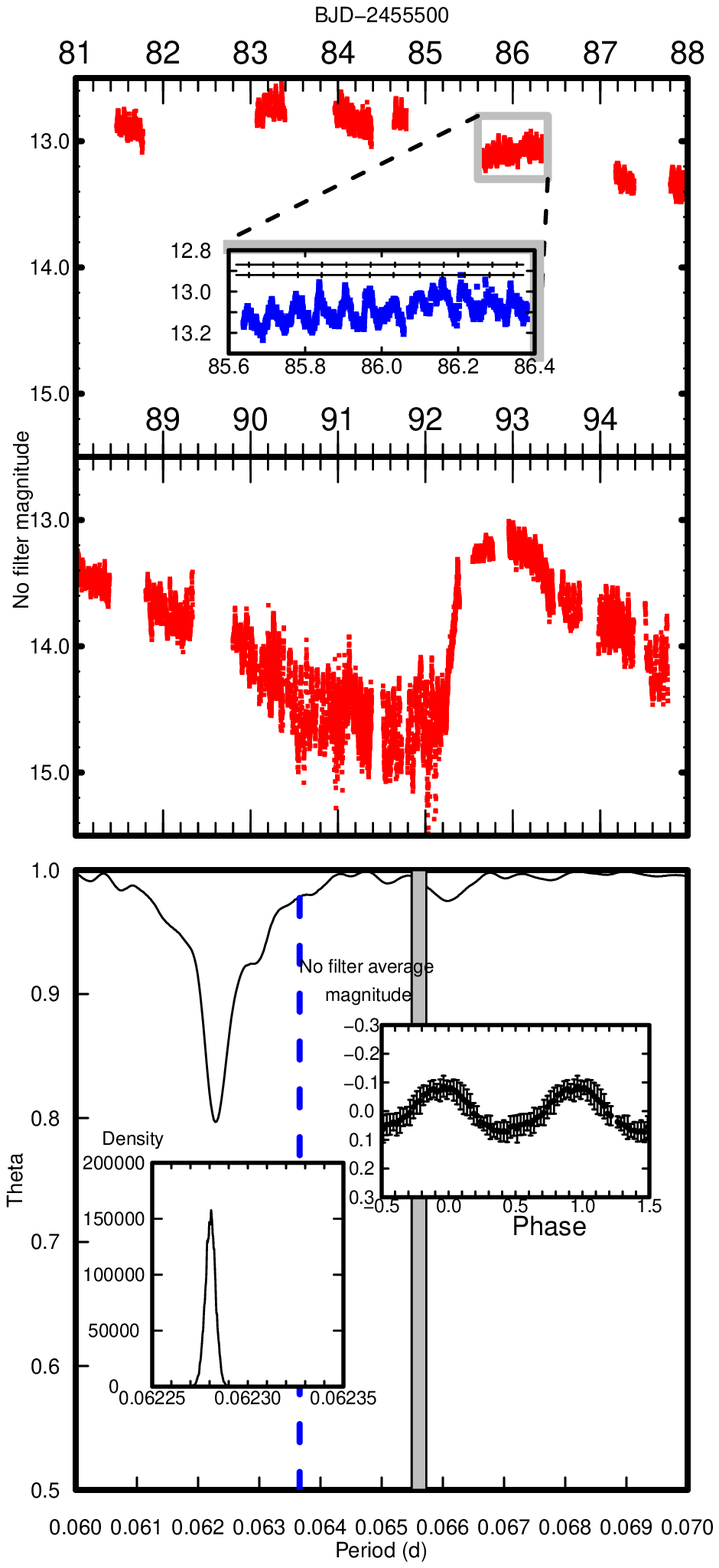}
\end{center}
\caption{Upper and middle panels represent the overall 
light curve of ER UMa superoutburst in
2011 January. The individual
points represent averaged observations in
0.02-date bins. The superoutburst stage at around 12.8 
mag is until BJD 2455591, and continuous
about a day is the quiescence (stayed at around 15 mag.). 
The next normal outburst started BJD 2455591 and reached 13 mag.

 The inset in the upper panel is a enlarged light
curve. The two lines with small ticks above the light curve 
are plotted at the interval of two kinds of period.
The ticks on lower line represent the timings of maxima of 
negative superhumps, and the ticks on upper line represent expected 
times assuming the period of orbital humps. The individual
points represent averaged observations in 0.0003-date bins.
The low panel represents PDM diagram. Theta stands for the
ratio of variance.
The blue dashed vertical line indicates the orbital period 
\citep{tho97erumav1159ori},
and the vertical gray strip indicates the range of the 
periods of positive superhumps
reported in earlier works \citep{kat95eruma}.
The left inset in the lower panel represents posterior
probability density function with the Markov-Chain Monte Carlo
analysis. The right one represents profile of negative superhumps 
phase-averaged by the period obtained by MCMC, 0.0622804 d
for the interval of BJD 2455587.8 -- 2455588.4.
Each point represents an average of each 0.025-phase bin. }\label{fig1}
\end{figure}

\begin{figure}[!htb]
\begin{center}
\FigureFile(80mm,50mm){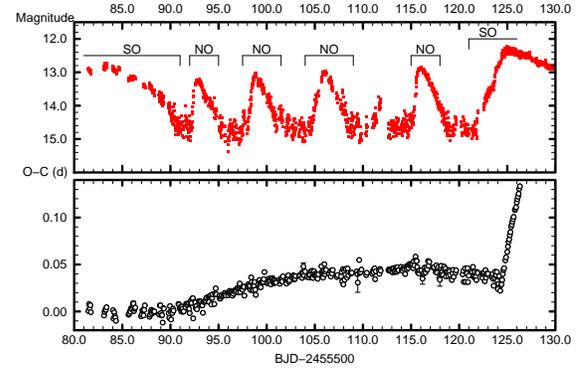}
\end{center}
\caption{The upper panel represents the overall light curve from
the superoutburst of 2011 January till the next superoutburst of
2011 March. SO represents for a superoutburst and NO represents
for a normal outburst. Although not observed by us, the superoutburst
of January is reported to have started on BJD 2455578 (January 16). 
The lower panel is $O-C$ diagram in the
same term. The negative superhumps had disappeared when 
the superoutburst
of March started and showed a smooth transition to positive superhumps.
As to circles without a error bar, errors are smaller than the circle 
radius (this applies to fig 3).
 }\label{fig2}
\end{figure}
\begin{figure}[!htb]
\begin{center}

\FigureFile(80mm,100mm){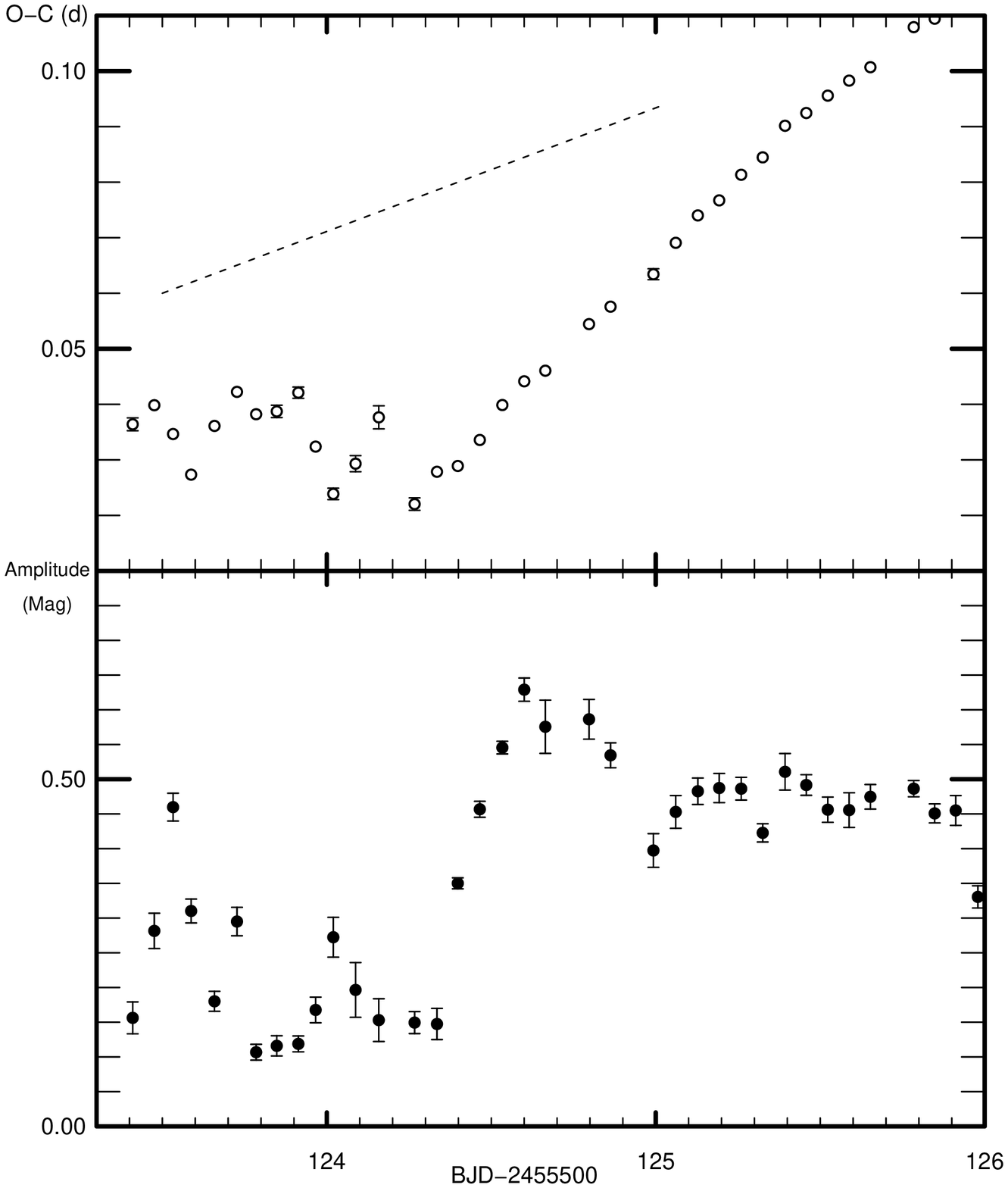}
\end{center}
\caption{The upper panel is enlarged $O-C$ diagram between 
BJD 245623.3 - 26.0.

This diagram apparently shows that the smooth transition from 
negative superhumps to positive superhumps without a phase shift.
The lower panel is variations of the amplitudes of superhumps.
The amplitude got to smaller shortly before this transition occurred.
The dotted line represents the $O-C$ variations in the case of variations
with the orbital period. This represents the transition from negative superhumps
to positive superhumps.
}\label{fig3}
\end{figure}

ER UMa underwent an outburst on 2011 January 16, and this outburst 
lasted for 12 days (fig \ref{fig1}.  Since this duration is far 
longer than those
(2-3 days) of ordinary (normal) outbursts of the same object,
and we identified this outburst as a superoutburst and conducted
a world-wide wide-band photometric campaign with CCD cameras 
equipped on 23.5 -- 40cm telescopes distributed on the globe.
The log of observations and instruments will be listed in 
forthcoming paper.
We performed dark subtractions and flat-fielding and measured the
differential magnitude against other stars using standard aperture
photometry.

 The overall light curve of the superoutburst is shown in upper
panel of fig \ref{fig2}.
Throughout these observations, large-amplitude variations up to 
0.6 mag were recorded (fig \ref{fig2}).

 We performed a period analysis with the Phase Dispersion Minimization
(PDM) method after zero-point adjustments between different observers and
the conversion to Barycentric Julian Date (BJD) and subtraction of the 
global trends of the outburst (interval BJD 2455581 - 89.5). 
This analysis yielded a unique period of 0.06226 d, safely excluding
any possibilities of either the orbital
period (0.06366 d, \cite{tho97erumav1159ori}) or positive
superhumps (0.06549--0.06573 d, \cite{kat95eruma}).  The 99\% confidence
limit of this period
was determined ($0.0622804^{+62}_{-66}$ d) by modeling the data using 
the Markov-Chain Monte Carlo
(MCMC) method and the averaged profile of variations \citep{kat09PDot2}.
These analysis clearly demonstrates the presence of a periodicity
shorter than the orbital period by 2.2\%.
There was no clear indication of positive superhumps evolving during this
superoutburst, contrary to all existing observations of superoutbursts.
The possibility that positive superhumps remained so weakly as not to appear
in these period analysis.
 
The times of maxima of negative superhumps
were determined by numerically fitting the light curve with the template
mean superhump profile for ER UMa obtained during this observation 
\citep{kat09}. A linear regression to these times 
yielded the following equation as the ephemeris of the superhump maxima:

\begin{equation}
\noindent
BJD(max) = 2455581.4540 (11) + 0.062242 (9) \times E
\label{ocequ}
\end{equation}

where E is the number of periodic cycles.

 The $O-C$ diagram using this equation is shown in fig \ref{fig2}.

 Negative superhumps did not disappear even after the termination
of the superoutburst,
in contrast to usual decay of positive superhumps.
The $O-C$ diagram indicates that the signals up to the next superoutburst
were essentially negative superhumps, with small systematic drifts in the
period (fig \ref{fig2}).
 Negative superhumps were also seen during four successive normal
outbursts and quiescence between normal outbursts. Negative superhumps
finally disappeared when the next superoutburst started on 2011 
March 3 and showed a smooth transition to ordinary superhumps
without a phase shift, when the amplitude of negative superhumps
got very small temporally (fig \ref{fig3}).

 Negative superhumps had been detected almost constantly although
negative superhumps were not observed clearly in the early phase 
of superoutburst. The behavior in longer-time scale will be described 
in the future paper.

\section{Discussion}

 The origin of negative superhumps
has not been well-understood.
 There is one attempt to explain the light modulations, observed
as negative superhumps, by considering the periodic variations of
luminosity of the stream impact point (hot spot) on a tilted
disk \citep{woo07negativeSH}.
If the disk is tilted, the stream can hit the outer edge of the accretion
disk only twice in per orbit, and the stream can impact the inner portion
of the disk at other times, and the cyclically variable gravitational
energy release produces negative superhumps.
This picture has an advantage in explaining that negative superhumps are
observed with high amplitudes in quiescence of dwarf novae,
when the luminosity of the hot spot dominates.
\citep{woo09} implied that negative superhumps were detected even in 
the case that
mass transfer shut off. However, in this paper, this phenomenon occurs 
when the mass ratio 
is rather large ($q \ge 0.30$) and the mass ratio of ER UMa system is not 
so large. Thus this effect will not be discussed in this paper.
 While the amplitudes of negative superhumps during the present outburst
reached 0.1--0.2 mag even in full outburst, a previous marginal detection
of negative superhumps during the rising stage of the same object reported
only 0.07 mag \citep{gao99erumaSH}.
The amplitude of negative superhumps in the present superoutburst is
thus far larger, and this amplitude is also far larger than the amplitude
of negative superhumps reproduced by the numerical simulation \citep{mon12NSH}.
This suggests that the present negative superhumps were more excited
than that in \citet{gao99erumaSH}. As suggested in \cite{mon09}, this
may be the change of the angle of tilt.
 
 The fact that negative superhumps were detected in the superoutburst
stage implies that another view. There is an argument that the slow 
growth rate of the dynamical tilt
instability requires sufficient time to grow, enabling them only
observable in long-lasting stable states as in novalike CVs 
\citep{osa95eruma}.
The present phenomenon would alternately implies that negative superhumps
can grow in much shorter time-scales under rapidly variable conditions.
There may be a mechanism in exciting negative superhumps other than
a dynamical tilt instability, and the mechanism may be related to the
one exciting positive superhumps. This view is also supported by the 
case of V1504 Cyg , where negative superhumps were
excited in failed superoutburst \citep{kat12Pdot3}.

It is noteworthy that the duration of the present superoutburst is
significantly shorter ($\sim$12 days) than those ($\sim$ 20 days) in usual
superoutbursts of the same object.  It is less likely this difference
was caused by a dramatic change in the global mass-transfer rate
since the interval (44 days) between the successive three superoutbursts,
which occurred on 2010 December 3, 2011 January 26, March 3, 
exactly matches the general supercycle, where the supercycle is
generally considered as a good measure of the global mass-transfer rate
\citep{osa89suuma}.

 It has been suggested that particularly long-lasting superoutbursts
in ER UMa and related systems are a result of an exceptionally large
mass-transfer rate from the secondary to the outer edge of
the accretion disk \citep{osa95eruma}.
 Assuming a tilted disk, the mass supply is prone to the inner portion
of the accretion disk, and the supply to the outer edge is expected
to be insufficient to achieve this condition.
The insufficient mass supply to the outer edge could naturally leads to
an early quenching of the superoutburst,
or prevents the excitation of the ordinary 3:1 tidal resonance
as observed in the present outburst.
Although there have been arguments whether long-lasting superoutbursts
are sustained by the 3:1 tidal resonance or by the irradiation-induced
enhanced mass-transfer, our present observations suggest that the
presence of the 3:1 tidal resonance producing positive superhumps
is more essential to maintaining long-duration superoutbursts.

Although the mechanisms to cause tilt are still unclear, 
\citet{monmar10} recently proposed that hydrodynamic lift
as the common source of disk tilts, and indicated that
there is a minimum mass-transfer rate to generate tilts.
The exceptionally high mass-transfer rates in ER UMa and related
systems \citep{osa96review}
may provide a favorable condition in continuously exciting tilts
even in various states, and this very condition might explain unusual
properties recorded in these objects (\cite{kat03erumaSH}; \cite{osa95rzlmi}).
The same high-mass transfer rate and potentially resultant triggered tilts
also universally explain occasional
low frequencies of normal outbursts, reported in dwarf novae
exhibiting negative superhumps(\cite{kat02v503cyg};
 \cite{kat02v344lyr}) and may be related to suggested 
early quenching of superoutbursts in some extreme 
dwarf novae(\cite{osa95rzlmi}; \cite{hel01eruma}).
the present discovery appears to be consistent with the
prediction by \citet{mon10} who suggested that negative superhumps
might appear in some high-dot{M} dwarf novae in outburst, including
SU UMa stars and ER UMa.

The interval of outbursts between the two superoutbursts
is $\sim$ 7 d, which is longer than one reported in 
\citet{rob95eruma} (4 d). 
This reduced number of outbursts also supports the interpretation
suppressing of normal outbursts.

The present
discovery of negative superhumps in an unprecedented condition
implies that the tilted or warped disk can be easily excited
in more universal conditions.
Tilted disks and precessing jets are universally, and in various
scales, seen or proposed in a variety of astrophysical objects
and jet systems, such as SS 433, X-ray binaries (ex: \cite{pri96};
\cite{monmar10}). Some of mechanisms suggested in these objects
require the strong radiation
field or the magnetism of the central object, which does not apply
to the present non-magnetic dwarf nova. This unexpected discovery
is expected to contribute to generally understanding the physics of
various fields of accretion disks. 

 The authors are grateful to observers of VSNET Collaboration who contributed
observations, covering both long-term variations and post-superoutburst
behavior of ER UMa.
 We are grateful to early detections and notifications of 2010 December
and 2011 January superoutbursts by Y. Maeda and G. Poyner, respectively.
 And we are grateful to Prof. Y. Osaki for
the discussion of the relation between negative superhumps and 
suppression of outbursts.
  This work is partly supported the Grants-in-Aid for the Global COE
Programme ``The Next Generation of Physics, Span from Universality 
and Emergence'' (G09) from the Ministry of Education, Culture, Sports,
Science and Technology (MEXT) of Japan.


\end{document}